\documentclass[useAMS,usenatbib]{mn2e}
\usepackage{graphicx}
\usepackage{url}
\usepackage{times}
\usepackage{aas_macros}
\usepackage[latin1]{inputenc}
\usepackage{times}
\usepackage{amssymb}

\title[Accretion rate and kinetic luminosity functions]{Measuring the accretion rate and kinetic luminosity functions of supermassive black holes}
\author[E. G. K\"ording, S. Jester,  R. Fender]{E. G. K\"ording$^{1}$\thanks{E-mail:
Elmar@phys.soton.ac.uk}, S. Jester$^{2}$, R. Fender$^{1}$\\
$^{1}$ School of Physics and Astronomy, University of Southampton
  Hampshire SO17 1BJ, United Kingdom\\
$^{2}$ Max-Planck-Institut f\"ur Astronomie, K\"onigstuhl
  17, 69117 Heidelberg, Germany
}

\newcommand{\Pj}{\ensuremath{P_{\mathrm{jet}}}}
\newcommand{\qj}{\ensuremath{q_{\mathrm{j}}}}
\newcommand{\Lbe}{\ensuremath{L_{\mathrm{bol}}^{\mathrm{equiv}}}}

\begin{document} 

\date{Accepted ? Received July 9, 2007 }

\pagerange{\pageref{firstpage}--\pageref{lastpage}} \pubyear{2007}

\maketitle

\label{firstpage}

\begin{abstract} 
We derive accretion rate functions (ARFs) and kinetic luminosity
functions (KLF) for jet-launching supermassive black holes. The
accretion rate as well as the kinetic power of an active galaxy is
estimated from the radio emission of the jet. For compact low-power
jets, we use the core radio emission while the jet power of high-power
radio-loud quasars is estimated using the extended low-frequency
emission to avoid beaming effects.  We find that at low luminosities
the ARF derived from the radio emission is in agreement with the
measured bolometric luminosity function (BLF) of AGN, i.e., all
low-luminosity AGN launch strong jets.  We present a simple model,
inspired by the analogy between X-ray binaries and AGN, that can
reproduce both the measured ARF of jet-emitting sources as well as the
BLF. The model suggests that the break in power law slope of the BLF
is due to the inefficient accretion of strongly sub-Eddington
sources. As our accretion measure is based on the jet power it also
allows us to calculate the KLF and therefore the total kinetic power
injected by jets into the ambient medium. We compare this with the
kinetic power output from SNRs and XRBs, and determine its
cosmological evolution.
\end{abstract}


\section{Introduction}
\label{s:intro}
There is increasing support for the idea that feedback from active
galactic nuclei (AGN) plays an important role for galaxy formation
\citep[e.g.,][]{CowieBinney1977,BinneyTabor1995,SilkRees1998,ChurazovKaiser2001,diMatteoSpringelHernquist2005}. This
feedback is also invoked to explain the M-$\sigma$ relation of
supermassive black holes
\citep[e.g.,][]{HaehneltNatarajanRees1998,King2003,RobertsonHernquistCox2006,FabianCelottiErlund2006}. It
is not yet clear whether kinetic or radiative feedback is dominant and
how efficient each is at a given accretion rate. In this paper we will
exploit the analogy between X-ray binaries (XRBs) and AGN to obtain
information about the efficiency of the different feedback processes
and calculate the total power available for feedback at a given
redshift.

The central engines of XRBs and AGN seem to be similar \citep[e.g.,][]{MirabelRodriguez1998,Meier2001} and
recently relations have been found which scale spectral and variability
properties from one class to the other
\citep{MerloniHeinzdiMatteo2003,FalckeKoerdingMarkoff2004,KoerdingJesterFender2006,McHardyKoerding2006}.
In the nearby universe ($z\la 0.2$), there are very few
high-luminosity quasars like those that exist at high redshifts. 
These bright high redshift quasars are likely to have a strong effect on 
the evolution of the galaxy luminosity function. However,
there are several XRBs that go through transient phases of very high
accretion rates. These ``very-high state'' objects may be better
templates for the bright quasars at a redshift of two than any nearby
AGN. Thus, we will use our knowledge of XRBs and their states to
obtain information about the kinetic and radiative properties of AGN.

For XRBs one can observe a full outburst cycle, in which the accretion
rate increases from very low ($10^{-7} \leq
\frac{\dot{M}}{\dot{M}_{Edd}} \leq 10^{-5}$) to near the Eddington
limit and then decays again. At low accretion rates, the source is
generally found in the \emph{hard state}, which is characterized by a
hard power law in the X-ray spectrum. The hard X-ray emission is
usually accompanied by radio emission associated with a compact jet
\citep{Fender2001} which sometimes can be directly imaged
\citep[e.g.,]{StirlingSpencer2001}.  The accretion flow in the hard
state is likely to be radiatively inefficient
\citep*[e.g.,][]{EsinMcClintockNarayan1997,KoerdingFenderMigliari2006}. The
source can stay in the hard state to fairly high luminosities ($\sim
30 \%$ Eddington) until it changes its state. During the state
transition, it first enters the \emph{hard intermediate state} (IMS),
which is characterized by a hard spectral component and band-limited
noise in the power spectrum. It is usually accompanied by an
increasingly unstable jet \citep{FenderBelloniGallo2004}.  After the
hard IMS it enters the soft IMS, which is dominated by a soft spectral
component and power-law noise in the power spectrum. Near this
transition one often observes a bright radio flare; however, after the
flare the jet seems to be quenched. After the soft IMS, the source is
often found in the \emph{soft state} where the X-ray spectrum is
dominated by a soft multi-colour black body component. Here the jet is
typically quenched by a factor $\ga 50$ in GHz radio luminosity
\citep{FenderCorbelTzioumis1999,CorbelFenderTzioumis2000}.  In the
soft state it is generally assumed that the source is efficiently
accreting, i.e., it has a standard geometrically thin, optically thick
accretion disk \citep{ShakuraSunyaev1973}. On the further way back
down to low accretion rates, the source stays in the soft state until
it reaches a critical accretion rate around $\sim 2\%$ of the
Eddington rate \citep{Maccarone2003}, where the reverse state
transition begins. This proceeds again via the soft IMS and the hard
IMS to the hard state. The transition from the hard to the soft state
typically occurs at a higher luminosity than the transition from the
soft to the hard state. Due to this hysteresis effect there is no
one-to-one correspondence of accretion rates to accretion states.  For
a  detailed discussion of states and their exact definitions see
\citet{Nowak1995,BelloniHomanCasella2005,HomanBelloni2005}, but see
also \citet{McClintockRemillard2003} for slightly different
definitions (mainly concerning the intermediate states).

Both in X-ray binaries and AGN, the unbeamed radio luminosity can be
used to estimate the accretion
rate. \citet{KoerdingFenderMigliari2006} do this for the \emph{core}
radio emission, while \citet{WillottRawlingsBlundell1999} present a
correlation between \emph{extended} radio luminosity and tracers of
the accretion rate of luminous double-lobed radio sources. Here, we
use these methods to translate radio luminosity functions into
accretion rate functions (ARF) of jet-emitting sources.  

In order to turn a \emph{radiative} bolometric luminosity function
(BLF) into an ARF as well, it is necessary to know the radiative efficiency of
the accretion flow. There have been several suggestions that
inefficient accretion is visible both in XRBs
\citep[e.g.,][]{Ichimaru1977,EsinMcClintockNarayan1997} and AGN (e.g.,
\citealt{ReesPhinneyBegelman1982,ChiabergeGilliMacchetto2006}), and a
number of authors have attempted to link both source classes and
establish a detailed correspondence (see, e.g.,
\citealt*{Meier2001,LivioPringleKing2003,MaccaroneGalloFender2003,MerloniHeinzdiMatteo2003,FalckeKoerdingMarkoff2004,Jester2005};
\citealt*{KoerdingFalckeCorbel2005};
\citealt{KoerdingFenderMigliari2006}). These sources are not only
likely to be inefficiently accreting, but the total energy output of
the sources may be dominated by the jet power
\citep[e.g.,][]{FenderGalloJonker2003,KoerdingFenderMigliari2006}.  To
compare the ARF with the BLF we will explore the effect of inefficient
accretion flows for low-luminosity objects that are included in the
bolometric luminosity function of
\citep{HopkinsRichardsHernquist2007}.  Comparing our radio-derived ARF
obtained in this way with this BLF, we can also determine a radio-loud
fraction that has an intuitive physical meaning: the ratio of volume
densities of radio-loud and radio-quiet sources at a given total
accretion rate.

As the jet power is employed in our method for estimating the
accretion rates from radio data, we can also directly estimate the
jets' kinetic power from the radio luminosities \citep[for an approach
  estimating jet powers from modeling blazar SEDs,
  see][]{MaraschiTavecchio2003}. This allows us to construct kinetic
luminosity functions for jet emitting sources in the local universe as
well as for higher redshifts. Such kinetic luminosity functions have
already been constructed by \citet{HeinzMerloniSchwab2007}. However,
we use a slightly different methodology and are able to include
high-luminosity sources (i.e., FR-II radio galaxies). By integrating
the kinetic luminosity functions over all luminosities we then
estimate the total power available for kinetic and radiative feedback
at a given redshift.  We also consider constraints on the Eddington
ratio distribution of accreting black holes which reproduce the
observed luminosity functions of \emph{all} classes of AGN using only
the black-hole mass function and prescriptions for the radiative and
kinetic outputs of accreting black holes as function of accretion rate
\citep[compare][]{VolonteriSalvaterraHaardt2006,MarulliBranchiniMoscardini2007}.

\section{Accretion rate and jet power estimates based on radio luminosities}

Low-luminosity AGN usually have a compact flat-spectrum radio core and
relatively weak optical emission
\citep{Ho1999,NagarFalckeWilson2005}. They probably have no ``big blue
bump'', and their total energy output of the source may be dominated
by the jet.  In the context of the analogy between XRBs and AGN these
objects are likely to correspond to hard-state objects, which exist
predominantly below accretion rates of $\sim 2 \%$ of the Eddington
rate. High-luminosity radio sources with significant extended emission
are typically classified as FR-II radio galaxies or radio-loud
quasars. These sources do have a big blue bump and are therefore
likely to be in the analogue of intermediate-state (IMS) sources. The
total power output of such sources is likely dominated by radiation.
As the accretion states depend on the Eddington-scaled accretion rate
(albeit with a hysteresis), a black hole that has a low mass but is
accreting strongly (e.g., an FR-II radio galaxy) may have a lower jet
power than a much more massive BH accreting in a strongly
sub-Eddington regime. Thus, simply referring to the two classes of
sources as ``low'' and ``high'' power sources is misleading.  We will
therefore refer to those sources accreting at a low fraction of the
Eddington rate and showing a compact radio core as ``LLAGN-like'' jet
sources. The radiation-dominated strongly accreting sources with
high-power jets will be referred to as ``RLQ-like'' jet sources.  This
division in two classes of jets is supported by similar differences in
XRB jets. Hard-state objects have relatively stable and compact jets
with a flat radio spectrum, while the IMS shows ejections of highly
relativistic blobs, i.e., an unstable jet.

\citet{KoerdingFenderMigliari2006} have presented a prescription for
estimating accretion rates of hard-state XRBs and unbeamed AGN from
the core radio luminosity.  Analysis of the VLBI monitoring of a
sample of flat-spectrum radio sources by \citet[the MOJAVE
  survey]{CohenListerHoman2006} has shown that only high-luminosity
sources have strongly beamed jets, while low-luminosity sources have a
lower maximum Lorentz factor, \emph{i.e.}, slower jets.  This is also
implied if the analogy between XRBs and AGN extends to jet speeds, as
\citet{GalloFenderPooley2003} found that hard-state jets are unlikely
to be strongly beamed while IMS jets may be faster \citep[but
  see][]{HeinzMerloni2004}.  Thus, low-power LLAGN-like AGN are
probably not affected strongly by beaming and we can use the
accretion-rate measure from \citet{KoerdingFenderMigliari2006} to
estimate their accretion rates from their core radio power, and hence
the accretion rate function (ARF) from their luminosity
function. We note that some FR-I RGs show high apparent
  velocities, in fact, they are thought to be the parent population of
  the highly beamed BL Lac objects. Nevertheless, they usually do not
  show a strong ``big blue bump'', so they may belong to the LLAGN-like jet sources.
 However, these FR-I RGs with high jet speeds seem to occur mainly at 
high radio luminosities \citep[$>10^{24}$W Hz$^{-1}$][]{{CohenListerHoman2006}} while we are mainly interested
at lower radio luminosities. Additionally, there do not seem to be any reported LLAGN blazars. 
While we cannot rule out that none of our LLAGN-like sources are beamed, the aforementioned arguments suggest
that our LLAGN sample will unlikely be strongly affected by beaming. 

However, at higher radio powers, it is currently not possible to
deboost a flat-spectrum core luminosity function reliably and convert
it into an ARF, since this would require the exact knowledge of the
distribution of Lorentz factors for a given accretion rate.  We
therefore estimate accretion rates using the core radio emission only
for low-power LLAGN-like jet sources.  For high-power RLQ-like jets,
we instead use the unbeamed extended low-frequency emission of the
lobes, whose strength has been shown to correlate with the
emission-line luminosity, and hence accretion rate, by the seminal
work of \citet{RawlingsSaunders1991}.

\subsection{Accretion rates and jet powers from core emission of
  LLAGN}

\subsubsection{Accretion rate}

\citet{KoerdingFenderMigliari2006} present an estimate of the
accretion rate of hard state XRBs and unbeamed AGN from the core radio
luminosity:
\begin{equation}
\dot{M} \approx 4 \times 10^{17} \left(\frac{L_{\mathrm{Rad}}}{10^{30}
  \mbox{erg s}^{-1}} \right)^{12/17} \frac{\mbox{g}}{\mbox{s}},
\end{equation}
This accretion measure is normalized using XRBs. As we extrapolate
this to AGN the uncertainty increases and we may underestimate the
fluxes slightly (see the mass term of the fundamental plane in
\citealt{KoerdingFenderMigliari2006}). Thus, as a consistency check we
will compare this accretion measure and jet power estimate to other
AGN estimates (sect.\,\ref{s:measure.compare} below).

\subsubsection{Jet power}

In addition to the accretion rate, one can also estimate the jet's
kinetic power form the core radio flux.  In fact, the accretion rate
is determined from the core radio power by assuming that the jet power is
linearly coupled to the accretion rate.  The fraction $\qj$ of the
accretion power which is injected into both jets $(\Pj = \qj\;
\eta\;\dot{M} c^2)$ is nearly a free parameter ($\qj \leq 1$). We now
determine this parameter by comparing our accretion measure with a
sample of sources with jet powers estimated independently of the core
power (see also \citealt{HeinzMerloniSchwab2007}).

The XRB Cyg X-1 has a well-measured jet power and core radio flux
\citep{GalloFenderKaiser2005,RussellFenderGallo2007}. We assume a
distance of 2.1\,kpc \citep{MasseyJohnsonDegioiaEastwood1995} and a
typical radio flux of 15\,mJy
\citep[e.g.,][]{GleissnerWilmsPooley2004}.  For the AGN, we use the
jet powers determined by \citet{AllenDunnFabian2006} and core radio
fluxes from the samples of
\citet{GiovanniniTaylorFeretti2005,NagarFalckeWilson2005}. All of
these jet power measurements are based on the power needed to inflate
an observed ``bubble''. They are therefore independent of our
accretion measure based on the radio core flux.

\begin{figure}
\resizebox{8.4cm}{!}{\includegraphics{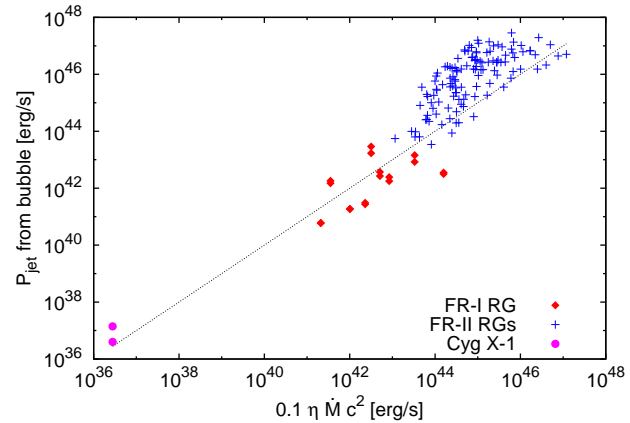}}
\caption{Jet power as a function of the accretion rate. We show the
  power injected into one jet as a function of the accretion rate as
  measured from the core radio emission. For Cyg X-1 and FR-I RGs we
  plot the jet power directly estimated from bubbles. For the FR-II
  RGs of the 3CRR sample we plot $0.1 \eta \dot{M} c^2$ for both
  methods to estimates the accretion rate.
\label{FigBubbles}}
\end{figure}
In Fig.~\ref{FigBubbles}, we show the jet powers determined for Cyg
X-1 and the \citeauthor{AllenDunnFabian2006} jets in comparison to our
accretion measure. The solid line assumes that the power in one jet is
$\approx 1/10$ of the available accretion power. Thus, if the kinetic
jet power estimates from bubbles are correct, we find $\qj \approx
0.2$ for the entire twin-jet system. This is consistent with the rough
estimate of the jet power given in \citet{KoerdingFenderMigliari2006}.
The total jet power can therefore be estimated:
\begin{equation}
P_{jet} \approx 7.2 \times 10^{36} \left(\frac{L_{\mathrm{Rad}}}{10^{30}
  \mbox{erg s}^{-1}} \right)^{12/17} \frac{\mbox{erg}}{\mbox{s}},
\end{equation}

Recently, \citet*{BinneyOmma2007} have reported simulations which
indicate that jet powers derived from the pressure and volume of jet
bubbles underestimate the true energy input by a factor of order 6. In
this case the jets would carry at least half of the power liberated in
the accretion disk.  The accretion disk could then radiate only half
of the accretion power. IF this is correct, all our accretion measure
estimates remain applicable, only the jet power and any kinetic
power estimate presented below has to be multiplied by a factor 6.

\subsection{Accretion rates and jet powers from extended radio emission}
\label{s:measure.extended}

\subsubsection{Accretion rates}
\label{s:measure.extended.acc}

The bulk of the radio luminosity of powerful radio galaxies is emitted by
extended lobes with unbeamed, steep-spectrum emission.
\citet{RawlingsSaunders1991} have shown that the strength of this
extended emission correlates with the narrow-line luminosity of the
optical core.  They used the correlation to argue for a common
mechanism powering both the radio jets and the optical continuum and
line emission. The reality of this correlation has been confirmed by
many later studies.  In particular,
\citet{WillottRawlingsBlundell1999} presented a refined analysis based
on complete radio surveys with vastly different flux limits, ruling
out distance effects as origin of the correlation.  They found a
correlation between rest-frame radio power at 151\,MHz and narrow-line
luminosity, and concluded that the jet power is within one order of
magnitude of the radiative power of the accretion disc.
As the disc luminosity directly measures the accretion rate, we can use
the correlation to determine an accretion rate from  151\,MHz radio
luminosities for high-power jet sources.

To allow a direct comparison of the ARF generated from these ``radio
lobe'' accretion rates with the ARF derived from the bolometric
luminosity of all quasars (\emph{i.e.}, including the majority of
quasars without strong radio lobes) by
\citet{HopkinsRichardsHernquist2007}, we do not directly use the
result of \citet{WillottRawlingsBlundell1999} but renormalize this
relation using $B$-band luminosities together with the bolometric
corrections of \citeauthor{HopkinsRichardsHernquist2007}.

We construct a sample of broad-line quasars with measured 74\,MHz
radio luminosities by cross-matching quasars from the SDSS DR5
\citep{SchneiderHallRichards2007} with the VLA Low-frequency Sky
Survey \citetext{VLSS\footnote{\url{http://lwa.nrl.navy.mil/VLSS/}},
  \citealp{CohenLaneKassim2006,CohenLaneCotton2007}} using a matching
radius of 20\arcsec\ (the histogram of radial separations between the
closest matched sources has a local minimum at this matching
radius). We use 74~MHz fluxes as this frequency is near the target of
151\,MHz, especially for the large number of the quasars around
$z\approx 2$, and as the VLSS provides an easily accessible deep
survey of the full northern sky.  The resulting matched list has a
total of 919 entries. We apply the $i$-band emission-line and
$K$-correction given in \citet[Table 4]{RichardsStraussFan2006} and
$K$-correct the VLSS data to 74~MHz rest-frame assuming a spectrum
$f_\nu \propto \nu^{-0.7}$.

\begin{figure}
\resizebox{8.4cm}{!}{\includegraphics{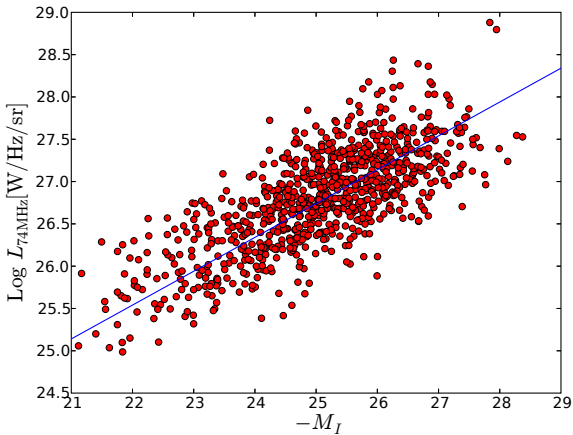}}
\caption{Normalizing the jet power measures from radio luminosities
  with sources of measured jet powers. On the ordinate ($y$-axis) we
  plot the 74 MHz~radio luminosity and on the abscissa ($x$-axis) we
  give the absolute $i$-band magnitude. The line represents the linear
  relation fitted to the data (see text). \label{FigL74}}
\end{figure}

In Fig.~\ref{FigL74} we show the 74 MHz luminosity against the
absolute $i$-band magnitude.  Following the nearly linear correlation
between the narrow-line luminosity and the low-frequency radio
luminosity found by \citet[see also
  \protect\citet{Willott2001}]{WillottRawlingsBlundell1999}, we fit a
linear dependence between the optical $i$-band luminosity (which we
assume is proportional to the ionizing luminosity, and hence to the
narrow-line luminosity) and the radio luminosity to the data:
\begin{equation}
\log L_{74} = -0.4 M_i + 16.78
\end{equation}

To convert the $i$-band magnitudes to $B$-band magnitudes, we assume a
power-law spectrum with $\alpha=-0.5$ \citep[see
][]{RichardsStraussFan2006} which yields $B-i = 0.3$. For the B-band
luminosity we find:
\begin{equation}
\log L_{\mathrm{B}} \left[\frac{\mbox{erg}}{\mbox{s}} \right] \approx \log
L_{74} \left[\frac{\mbox{W}}{\mbox{Hz sr}} \right] + 18.6.
\end{equation}
Throughout this paper we mainly use cgs units. However, as most radio
luminosity functions are given in W\,Hz$^{-1}$\,sr$^{-1}$ we provide
the conversion formulae from radio luminosity to accretion rate using
these units on the right hand side of the equation.
\citet{HopkinsRichardsHernquist2007} use a non-constant bolometric
correction of the $B$-band flux. However, the bolometric correction
deviates from a constant mainly at lower luminosities where we will
not use the relation to obtain accretion rates. As we would like to
obtain a simple relation between the low frequency radio luminosity
and the bolometric luminosity we will use a constant bolometric
correction of 10, i.e., $L_{bol} \approx 10 L_B$. Within our
uncertainties, this is in agreement with the value used by
\citet{HopkinsRichardsHernquist2007}. If we again assume that radio
emission has a spectrum $f_\nu \propto \nu^{-0.7}$, we can translate
the measured relation to:
\begin{equation}
\log L_{\mathrm{bol}} \left[\frac{\mbox{erg}}{\mbox{s}} \right] \approx \log
L_{151} \left[\frac{\mbox{W}}{\mbox{Hz sr}} \right] + 19.8 . \label{extendedeq}
\end{equation}
To obtain the accretion rates we assume that the sources are accreting
efficiently with a constant efficiency of $\eta = 0.1$:
\begin{equation}
\log \dot{M} \left[\frac{\mbox{g}}{\mbox{s}}\right]  = \log L_{151} \left[\frac{\mbox{W}}{\mbox{Hz sr}} \right] - 0.15  \label{Mdotform151}
\end{equation}

The correlation between optical narrow-line luminosity and 151 MHz
radio luminosity \citep{WillottRawlingsBlundell1999} has only been
tested with a sample in the luminosity range of $10^{24.5} \leq L_{\rm
  151 MHz} / (\mbox{W Hz}^{-1} \mbox{sr}^{-1}) \leq
10^{28.8}$. Therefore, we will only use it for extended radio emission
brighter than $10^{24.5}$ W\,Hz$^{-1}$. 

\subsubsection{Jet power}
\label{s:measure.extended.jetpower}

If we assume that the ratio of jet power to accretion rate $\qj$ is
similar for hard-state and IMS objects, $\qj \approx 0.2$, we can use
the accretion rate determined from the extended low-frequency radio
emission also as a measure of the jet power.
\citet{WillottRawlingsBlundell1999} and
\citet{HardcastleEvansCroston2007} report that
\begin{equation}
\Pj \left[\frac{\mbox{erg}}{\mbox{s}} \right] \approx 3 \times10^{17} L_{151}\left[\frac{\mbox{W}}{\mbox{Hz sr}} \right]  f^{3/2},
\end{equation}
where $f$ parameterizes our uncertainty of the jet power compared to
the minimum energy needed to account for the synchrotron emission from
the lobes.  \citet{BlundellRawlings2000} suggest that $f\approx 10$ is
applicable to FR-II RGs. We find:
\begin{equation}
\log \Pj  \left[\frac{\mbox{erg}}{\mbox{s}} \right]\approx 18.7 + 
 \log L_{151} \left[ \frac{\mbox{W}}{\mbox{Hz sr}} \right] .
\end{equation}
Our normalization of the $L_{151}$-accretion rate correlation
(eq.~\ref{extendedeq}) together with $\qj=0.2$ gives a normalization
constant of 19.1 compared to 18.7 as estimated from
\citet{WillottRawlingsBlundell1999}. The difference corresponds to a
factor 2.5. This is well within the uncertainties of our accretion
rate and jet power estimates, but may indicate that our normalization
is slightly too high. 

For the rest of this paper we will assume that the coupling constant
of the jet is $\qj=0.2$ and use this to estimate the jet power from
the accretion rate for both low-luminosity and high-luminosity
objects. Thus, our jet power measure from extended radio emission is:
\begin{equation}
\log \Pj  \left[\frac{\mbox{erg}}{\mbox{s}} \right]\approx 19.1 +  \log L_{151}\left[\frac{\mbox{W}}{\mbox{Hz sr}} \right].
\end{equation}

\subsection{Comparison of jet and accretion rate measures from both
  methods}
\label{s:measure.compare}

In the preceding subsections we have presented two different accretion
rate and jet power measures based on radio luminosities. As we will
use these to obtain accretion rate functions, it is important that the
estimates are consistent with each other.

To compare the accretion rate measures, we need a sample which has
measured values both for the extended low-frequency radio flux and for
the unresolved core flux, so that we can compute and compare both
accretion rate measures for the same sample.  The FR-II radio galaxy
subsample of the 3CRR catalogue fulfills these criteria. We take
151\,MHz radio fluxes from \citet{LaingRileyLongair1983} and core
radio fluxes from the compilation by
M.\ Hardcastle\footnote{\url{http://www.3crr.dyndns.org}}.  Since the
core fluxes of this sample are probably affected by beaming, we can at
the same time assess the impact of beaming.

In Fig.~\ref{FigBubbles} we show the 3CRR FR-II sample together with
Cyg X-1 and the \citet{AllenDunnFabian2006} sample of sources with
have jet powers inferred from X-ray bubbles. Most of the 3CRR sample
lies above the expected line, \emph{i.e.}, the jet power estimated
from the extended emission exceeds that estimated from core radio
fluxes. While the core flux is affected by beaming, the extended
emission is \emph{not}.  For a randomly orientated sample of sources,
like the low-frequency selected 3CRR sample, the majority of sources
will be \emph{de}boosted, since they are beamed into a narrow cone
(for $\Gamma \gtrsim 1.7$).  This explains the lower average jet power
estimated from the core.

An additional problem that may alter the relative distribution of the
two accretion power estimates is that the extended emission of the
lobes can only yield a measure of the jet power averaged over the
lifetime of the lobes (several million years). In contrast, the core
radio fluxes measures the instantaneous jet power. Thus, if a source
changes from having a strong jet to having a quenched jet, but the
relativistic particles in the lobes keep radiating for at least a
synchrotron lifetime, this will reduce only the core power
measure. The reverse effect, i.e., a starting jet, can also
happen. However, as the 3CRR sample is selected using low-frequency
radio emission those starting sources will not be in the sample.

\begin{figure}
\resizebox{8.4cm}{!}{\includegraphics{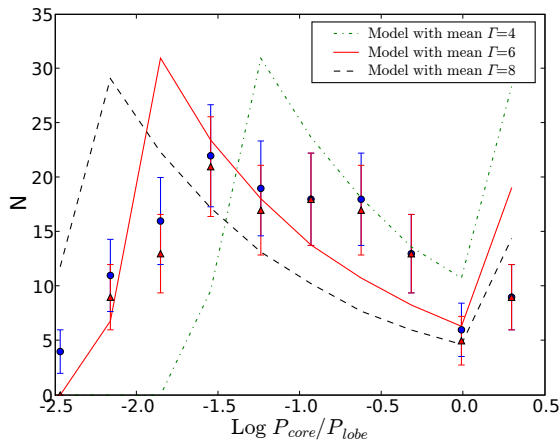}}
\caption{Beaming Statistics of the core radio measure compared to the
  accretion measure based on the extended emission: On the $x$-axis we
  show $\log P_{\mathrm{jet,core}}/P_{\mathrm{jet,lobe}}$ while the
  $y$-axis denotes the number of sources. The filled circles represent
  the measured histogram including upper limits and the triangles show
  only those sources with measured core fluxes. The lines describe a
  beaming model assuming a Gaussian distribution of Lorentz factors
  around $\Gamma = 4,6,8$ with $\sigma = 1$. \label{FigBeamingStats}}
\end{figure}
To verify that the ratios between the core jet power estimate and the
lobe jet power estimates are consistent, we show a histogram of the
measured ratios in Fig.~\ref{FigBeamingStats}. Some of the sources in
the 3CRR catalogue have no measured core radio fluxes. We therefore
present two histograms: In the first, we only include detected
sources. In the second we include all non-detected sources at their
detection limit. In addition to the measured distribution we also
present the distribution expected from a randomly orientated sample of
relativistic jets. Here, we assume that the observed luminosity is
relativistically boosted as $ L \sim D^{2.7} L_0$, where $L_0$ is the
luminosity in the rest-frame of the jet and $D$ is the Doppler factor
$D = 1/[\gamma (1-\beta cos \theta)]$.  We have assumed that the
Lorentz factors are Gauss-distributed around a Lorentz factor
$\Gamma_0$ with $\sigma = 1$. We show the model for three different
mean Lorentz factors: $\Gamma = 4,6,8$. While the model assuming
$\Gamma = 4$ over-predicts the number of boosted objects
($P_{lobe}/P_{core} \geq 1$) it under-predicts the number of strongly
deboosted sources. The model assuming $\Gamma = 8$ strongly
over-predicts the number of deboosted sources. The data is best
represented by the model with $\Gamma=6$. This is consistent with the
average Doppler factors found in BL Lac objects \citep[$D \sim
  7$;][]{GhiselliniPadovaniCelotti1993}.  The deviations can be
attributed to the simplicity of the model, as well as to short-term
changes in the accretion rate which only show in the accretion measure
based on the core fluxes. We conclude that both accretion rate
measures yield results which are in agreement with each other.

\section{Results}

We have constructed accretion and jet power measures from radio
luminosities. We now use these to translate radio luminosity functions
into accretion rate functions and compare those to the measured
bolometric luminosity function of quasars (BLF) from
\citet{HopkinsRichardsHernquist2007}.  For low-luminosity objects we
use the local radio luminosity function of
\citet{FilhoBarthelHo2006}. The high-luminosity end is obtained from
the low-frequency radio luminosity function of
\citet{WillottRawlingsBlundell2001}.

\subsection{Local accretion rate functions} \label{selocallum}

\begin{figure}
\resizebox{8.4cm}{!}{\includegraphics{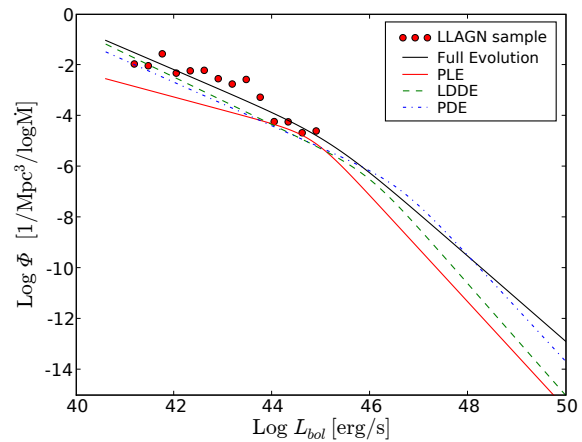}}
\caption{Different models for the cosmological evolution of the
  bolometric luminosity function extrapolated to $z=0$. The points
  represent the measured LLAGN radio luminosity function translated to
  accretion rates. To allow for an easy comparison with the
  bolometric luminosity function we show the accretion rates as an
  equivalent bolometric luminosity (0.1 $\dot{M} c^2$).  To present
  the uncertainties due to the model of the cosmological evolution we
  show four different models: Full evolution, PLE, PDE and LDDE. Only
   the latter model is used in the remainder of this
  paper. \label{FigLumZ0} }
\end{figure}
In Fig.~\ref{FigLumZ0} we show the local ARF of LLAGN-like jet sources
obtained from our accretion rate measure together with several models
of the BLF in the local universe (extrapolated to $z=0$,
\citealt{HopkinsRichardsHernquist2007}).  To compare the ARF with the
BLF, we define an equivalent bolometric luminosity $\Lbe = 0.1 \dot{M}
c^2$, \emph{i.e.}, we plot the ARF as if the sources were accreting
efficiently, with an efficiency of converting rest-mass energy into
radiation of $\eta = 0.1$. If the majority of the sources in the BLF
are indeed accreting efficiently, we can compare the ARF and the BLF
directly in this way. This assumption is likely to be true for the
high-luminosity quasars. We will discuss the effects of inefficient
accretion on the shape of the BLF in section \ref{s:disc.break}.

Of the models given in \citet{HopkinsRichardsHernquist2007}, we show
the pure luminosity evolution (PLE) model, the pure density evolution
(PDE) model, the luminosity dependent density evolution (LDDE) model
and the full evolution model. The latter is basically a broken power
law model where all parameters (both slopes and the break location)
can have a different evolution (for the exact definitions of the
different models see the original paper). We show the different models
mainly to demonstrate the uncertainties of the modeling and will adopt
the LDDE model for the rest of this paper as it currently seems to be
the most widely used model and the 151~MHz luminosity function is
fitted with an LDDE type model.  The BLF has data reaching to
$10^{42.1}$ erg/s, below that the models are extrapolated.
\begin{figure}
\resizebox{8.4cm}{!}{\includegraphics{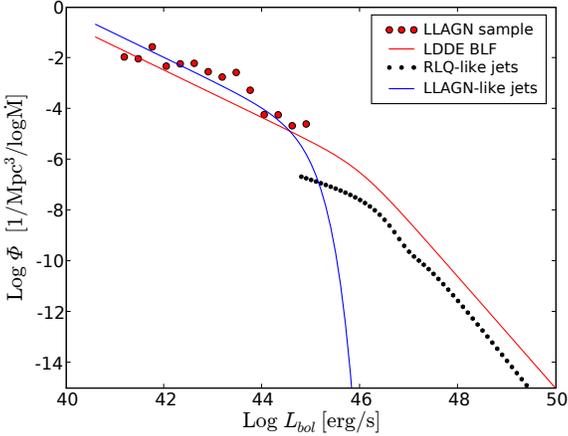}}
\caption{Bolometric luminosity function at $z=0$ \label{SlowFast}; we also  show the measured 
  ARF of ``low''-power jet sources (LLAGN-sample) and RLQ-like jet sources (e.g., radio-loud quasars). The latter ARF is obtained from a 151~MHz luminosity function together with eq.~(\ref{Mdotform151}).
The fit to the ``low''-power jets is the model
  of the BLF modified to include an exponential cut-off as seen in the
  black hole mass function -- see text.  }
\end{figure}

Our ARF obtained from the local radio luminosity function is in
agreement with the BLF, within the uncertainties of the cosmological
evolution. Thus, at low accretion rates there are similar numbers of
radio jet-emitting sources and sources appearing in the BLF of
\citeauthor{HopkinsRichardsHernquist2007}. 

Our accretion measure using core radio fluxes is strongly affected by
relativistic beaming. However, no AGN below a radio luminosity of
$\approx 5 \times 10^{22}$ W/Hz shows high apparent velocities
\citep{CohenListerHoman2006}. This radio luminosity corresponds to a
bolometric luminosity of $\approx 2 \times 10^{44}$ erg/s. This
supports the idea that the low-luminosity radio LF is not strongly
affected by relativistic beaming. We mentioned that we expect LLAGN-like
jets to exist predominantly below $\sim 2 \%$ of the Eddington
rate. For a $10^8$ M$_\odot$ BH this corresponds to a bolometric 
luminosity of $\sim 4 \times 10^{44}$ erg s$^{-1}$. Thus, LLAGN-like jet
are likely to be slower than RLQ-like jets.

The majority of the sources in the low-luminosity radio LF would not
be classified as Quasars as they are faint in the optical band and/or
do not show broad emission lines. Even though the BLF of
\citet{HopkinsRichardsHernquist2007} is called a ``Quasar LF'', it
also contains lower-luminosity AGN in addition to Quasars. This partly
explains why the ARF of LLAGN-like jets extends the BLF smoothly to
lower accretion rates. This will be further discussed in
Sect.~\ref{s:disc.break}.

\begin{figure*}
\resizebox{8.4cm}{!}{\includegraphics{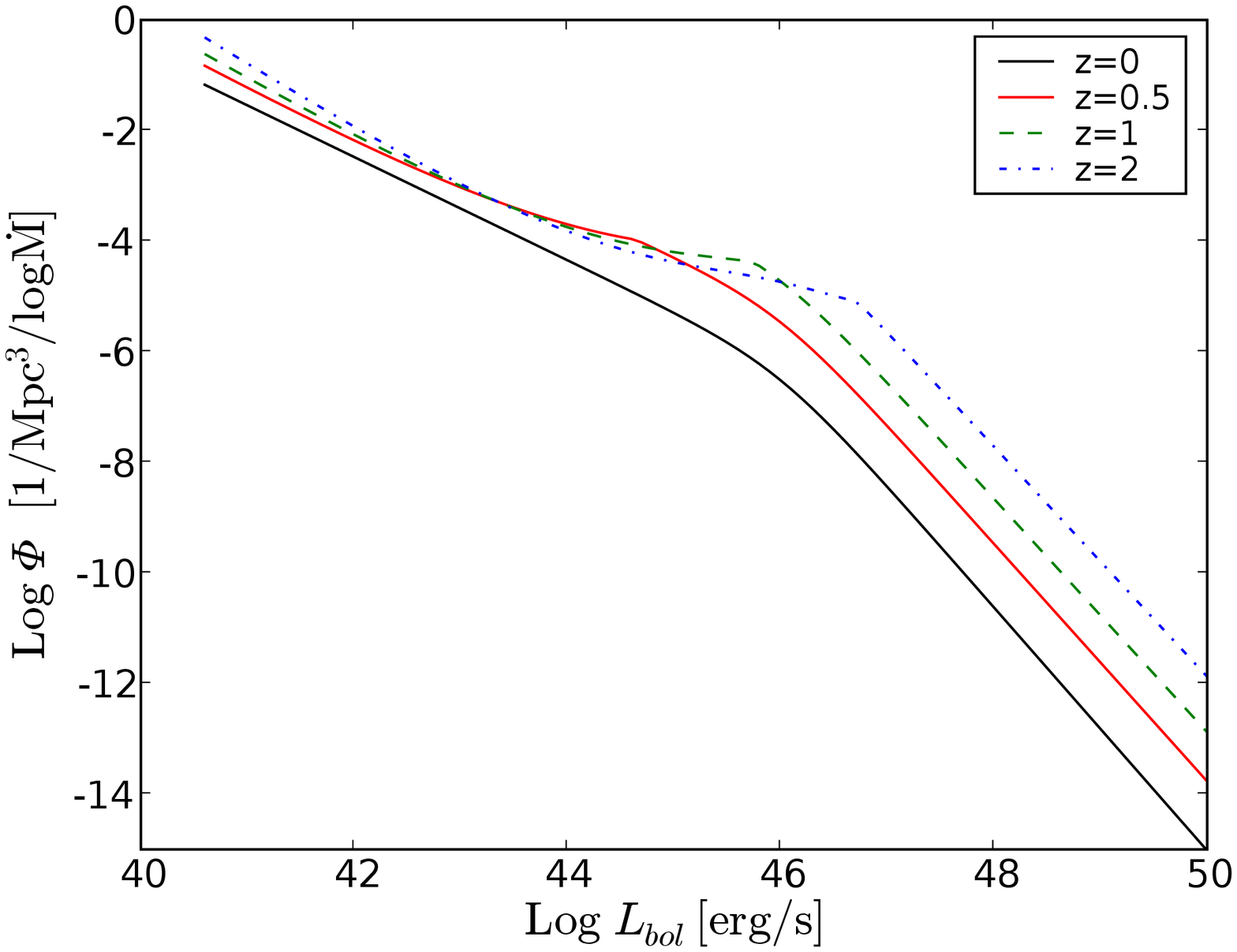}}
\resizebox{8.4cm}{!}{\includegraphics{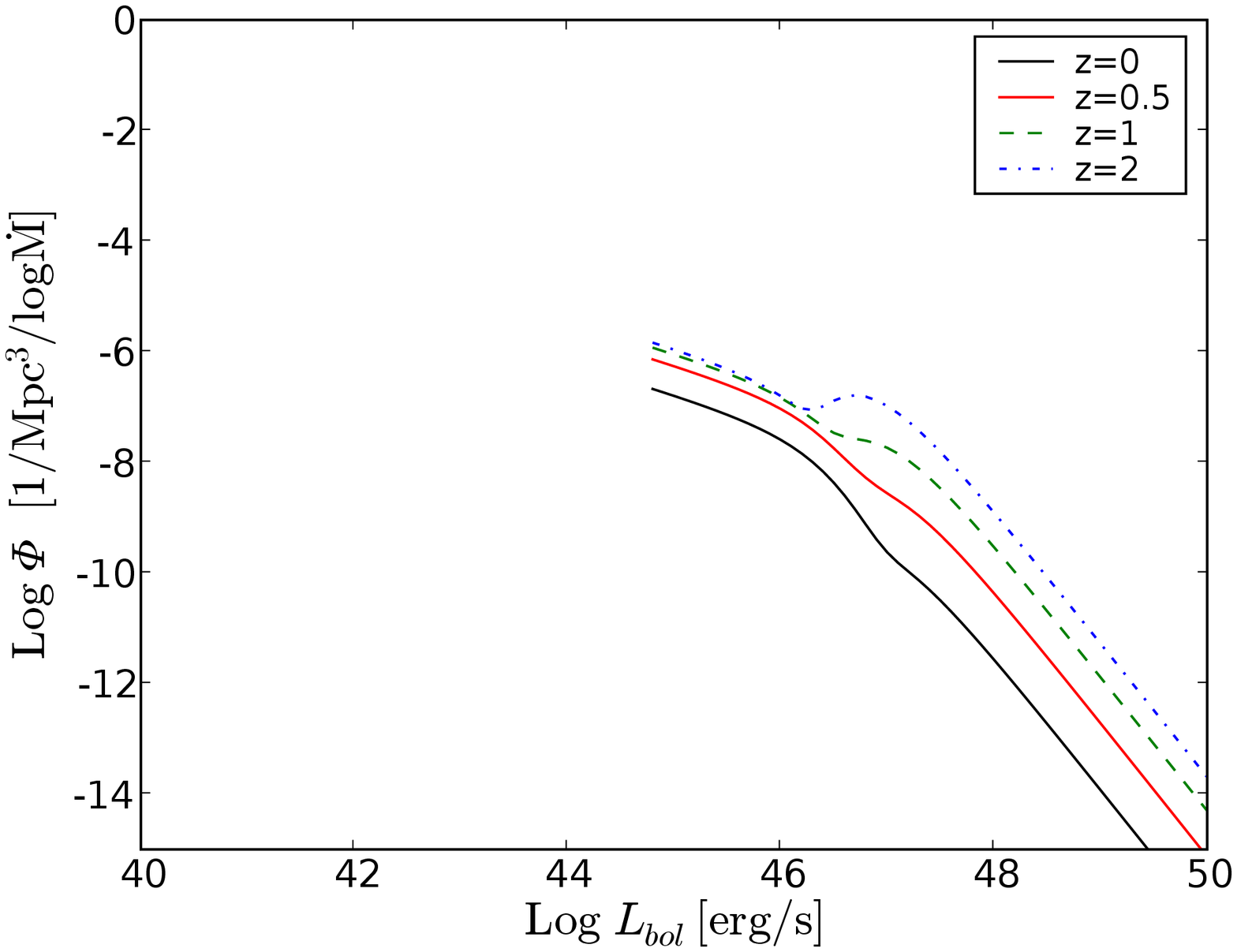}}
\caption{Luminosity function at different redshifts: In the left panel
  we show the evolution of the BLF at four different redshifts (0,0.5,1,2). This
  is a reproduction of the LDDE model given in
  \protect\citet{HopkinsRichardsHernquist2007} to allow for a direct
  comparison with the ARF of RLQ-like jets. In the right panel we show
  the evolution of the ARF of RLQ-like jet sources. The ARF shown is
  based on the 151 MHz radio LF of
  \protect\citet{WillottRawlingsBlundell2001}. \label{Figzevo}}
\end{figure*}
In Fig.~\ref{SlowFast} we show the measured ARF of the LLAGN-like jet
sources together with the BLF and the ARF deduced from the 151 MHz
radio luminosity function. The latter ARF describes the abundance of
radio-loud Quasars with lobes, \emph{i.e.}, radio-quiet quasars will
not be picked up. While jets of LLAGN-like sources do not seem to be
highly relativistic, those of RLQ can be highly relativistic. We will
refer to the ARF deduced from the 151 MHz LF also as \emph{the
  luminosity function of ``RLQ-like'' jets}, and we will use the term
``strongly jet emitting sources'' to refer to \emph{all} jet-emitting
sources, whether the jet is RLQ-like or LLAGN-like.  Sources without
strong jets will be referred to as \emph{radio-quiet}.

While the LLAGN-like jet ARF is roughly in agreement with the BLF, the
RLQ-like jet ARF lies significantly below it.  However, continuity
demands that the overall ARF of all jet-emitting sources connects the
RLQ-like jet LF smoothly to the LLAGN-like jet ARF.  The exact shape
of the ARF between the observed domains of LLAGN-like jets and
RLQ-like jets is unknown.  We attribute the LLAGN-like jet sources to
the hard state, which exists predominantly below an accretion rate of
$\sim 2 \%$ Eddington.  For a given accretion rate only black holes
above a corresponding minimum mass will have an Eddington ratio below
$\sim 2 \%$ and always have LLAGN-like jets. Above that Eddington
ratio it is likely that the majority of AGN will either show RLQ-like
jets or be radio-quiet.  Therefore, we can obtain constraints on the
ARF of LLAGN-like jet sources from the mass function of supermassive
black holes in the local universe.

\citet{ShankarSalucciGranato2004} report that the mass
function can be described by a power law with an exponential cutoff:
\begin{equation}
\Phi_{\mathrm{BH}}(M) = 7.7\times 10^{-3}
\left(\frac{M}{M_*}\right)^{-0.11}
\exp\left[-\left(\frac{M}{M_*}\right)^{0.45}\right], \label{eqBHfunc}
\end{equation}
with $M_* = 6.4 \times 10^{7} M_\odot$.  The BH mass function
describes how many BHs can harbour LLAGN-like jets for any given
accretion rate. The measured LLAGN-like jet ARF can be well described
by a power law -- similar to the BH mass function at low BH masses
(albeit with a different power law index). As the BH mass function has
an exponential cut-off the power law describing the LLAGN-like jet ARF
has to cut-off at some point as well. If the majority of AGN with an
Eddington ratio larger than $\sim 2 \%$ does not show LLAGN-like jets,
the BH mass function suggests that the cut-off of the ARF will be the
same exponential translated to a luminosity corresponding to $\sim 2
\%$ Eddington for BHs with mass $M_*$ (this luminosity is
$\sim 10^{44.2}$ erg/s).  As the LLAGN-like jet ARF is roughly in
agreement with the BLF, we therefore describe the LLAGN-like jet ARF
with the analytic model of the BLF with an additional exponential
cutoff and a constant logarithmic offset.  This model with an offset
of 0.5 dex is shown in Fig.~\ref{SlowFast}. The total ARF of
jet-emitting sources is the sum of both radio ARFs derived from
RLQ-like jets and LLAGN-like jets, respectively.

We can now revisit our assumption that the low-luminosity AGN have
slow jets. If this assumption was wrong, we would have to correct the
luminosity function for beaming. However, we are mainly interested in
the low-luminosity end of the radio luminosity function. At the lowest
luminosities the beaming correction would in fact increase the total
number of sources \citep[see e.g.,][]{Lister2003}. However, the ARF
already gives a density of $\sim 10^{-2}$ BHs per Mpc$^3$ and $\log
\dot{M}$ for accretion rates below an equivalent bolometric luminosity
of $\sim 10^{43}$ erg s$^{-1}$. If the beaming correction increased this
value significantly (e.g., a factor 10) this would suggest that there
were more low-luminosity AGN than BHs in the local universe. Thus,
even if there is a small beamed contribution to the luminosity of
LLAGN, the beaming correction cannot be strong.

\subsection{Redshift evolution of accretion rate functions}

In Fig.~\ref{Figzevo} we show the redshift evolution of the BLF
(left-hand panel) and the evolution of the ARF of RLQ-like jet objects
(right-hand panel). As the LLAGN-like jet ARF at redshift zero is
roughly in agreement with the BLF, we assume that this is also the
case for other redshifts. In this case, its cosmological evolution is
identical to that of the low-luminosity end of the BLF and we do not
show it separately.
\begin{figure}
\resizebox{8.4cm}{!}{\includegraphics{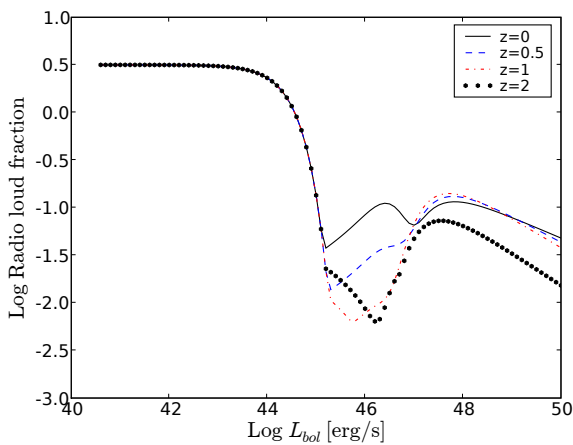}}
\caption{Fraction of radio-loud objects (\emph{i.e.}, volume ratio of
  radio-loud objects to the BLF) as function of accretion rate for
  various redshifts (0,0.5,1,2). At low luminosities all sources are likely to be
  radio-loud.  No evolution is visible, as we assume that the
  radio-loud LLAGN extend the BLF at low luminosities. The evolution
  at high luminosities is low for moderate redshifts ($z<2$). 
  Only at intermediate luminosities there seems to be a small
  region showing some evolution. However, this is likely an artifact
  of the choice of our models. Our accretion rate measures work best
  at the lowest and highest luminosities.  \label{Figfrac}}
\end{figure}

In Fig.~\ref{Figfrac} we show the fraction of strongly jet emitting
objects for a given accretion rate per volume. The number of strong
jet sources is the sum of the number of the LLAGN-like and the
RLQ-like jets. At low accretion rates, nearly all sources are
radio-loud. The radio ARF is in agreement with the low-luminosity end
of the BLF, or lies slightly above it. At the high-luminosity end, the
number of RLQ-like jet sources of a given accretion rate is
significantly lower than the number of quasars ($\sim 1$ dex). The
exact fraction of jet emitting sources is strongly dependent on the
normalization of the accretion rate measure based on the extended
radio emission. The slope of the BLF is $\sim -2.5$. Thus, all
uncertainties in the estimation of the accretion rate enter at least
quadratically in the radio-loud fraction.  We found in
Sect.~\ref{s:measure.extended.jetpower} that our jet power estimates
are slightly higher than those of
\citet{WillottRawlingsBlundell1999}. If we overestimate the jet power
by 0.2 dex, the true fraction of radio-loud objects at high
luminosities is less than $3\%$, well in agreement with the values
typically found.
 \begin{figure*}
\resizebox{8.4cm}{!}{\includegraphics{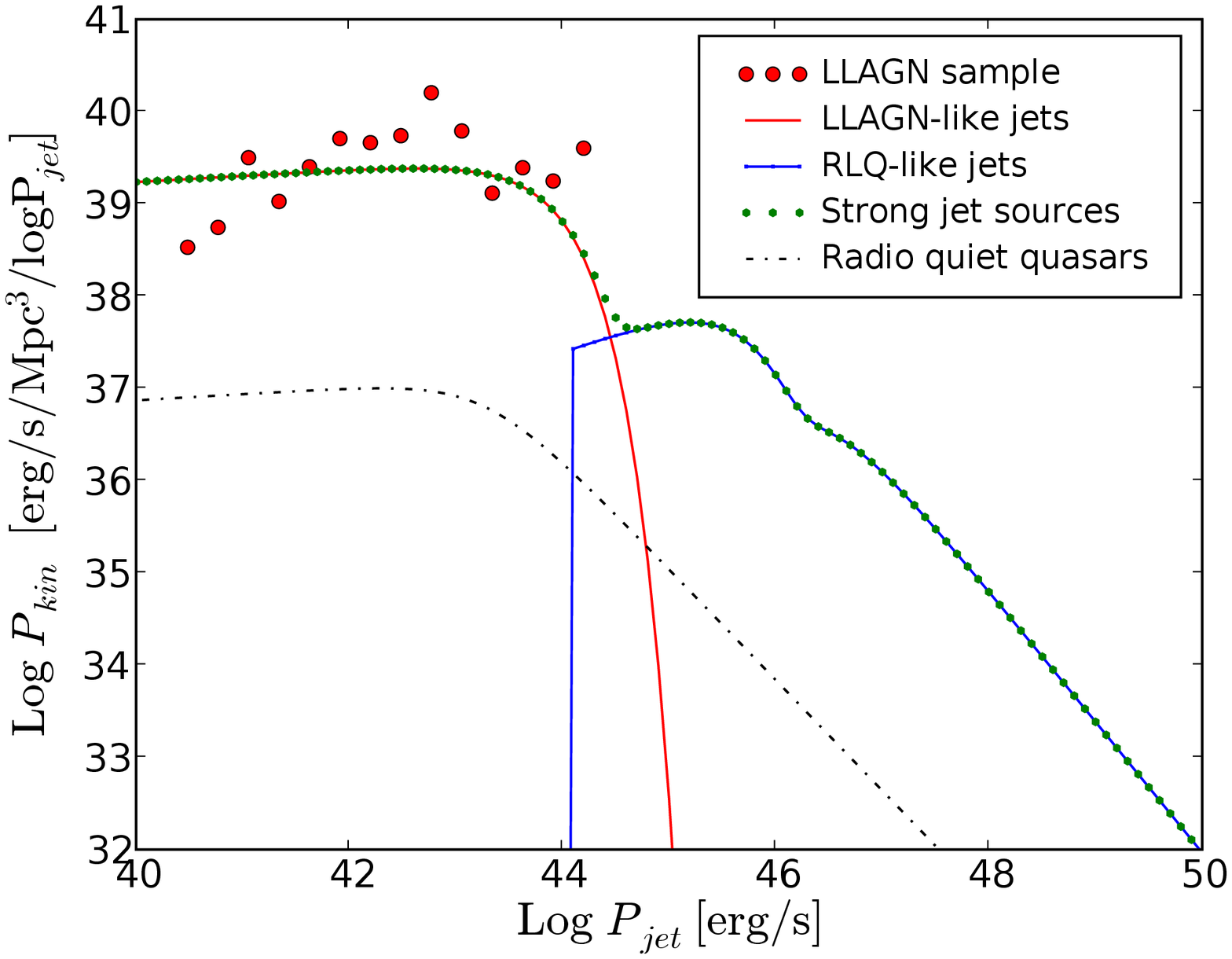}}
\resizebox{8.4cm}{!}{\includegraphics{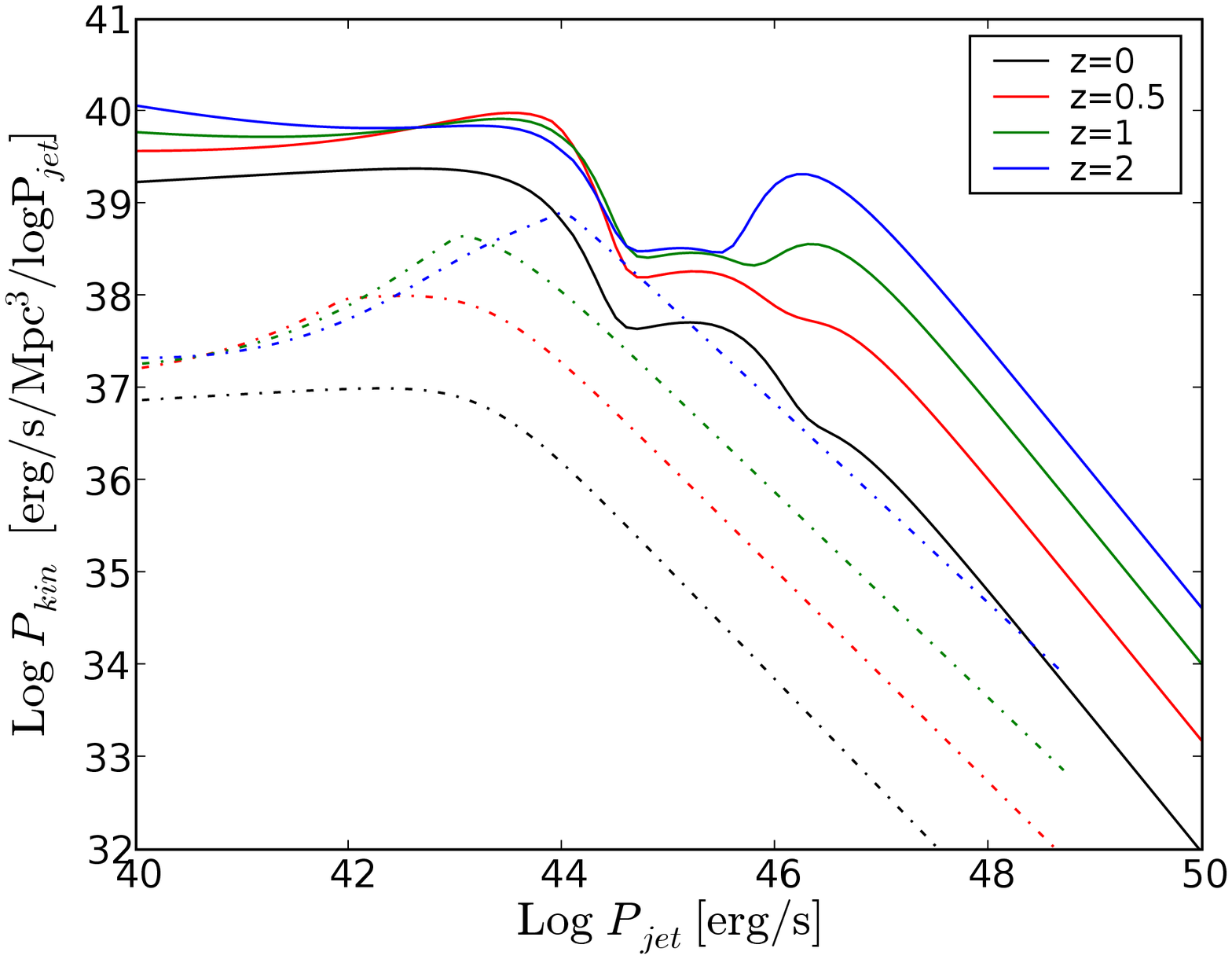}}
\caption{The kinetic luminosity function of jets shows the kinetic
  power injected into the ISM/IGM per Mpc$^3$ and log$_{10} \Pj$ as a
  function of jet power $\Pj$. The left-hand panel shows the different
  components of the kinetic luminosity function at redshift zero: LLAGN-like jet sources, RLQ-like sources as well as radio-quiet sources.
 The right-hand panel shows the evolution of the kinetic luminosity
  function of jet sources (RLQ-like and LLAGN-like jets, solid lines)
  and radio-quiet sources (dashed lines).
\label{FigLJet}}
\end{figure*}
\subsection{Jet kinetic luminosity functions} \label{sekinfeed}

As our accretion measures from radio fluxes are based on the radio
emission from the jet, they can also be used as a measure of the jet
power (see Fig.~\ref{FigBubbles}). Thus, we can compute the kinetic
luminosity function (KLF) for jets, \emph{i.e.}, measure how much
power is injected into the interstellar or intergalactic medium
(ISM/IGM) by jets from sources with a given accretion rate. In
sect.~\ref{s:measure.extended.jetpower}, we found that roughly a
fraction $\qj=0.2$ of the accretion power is injected into the jet for
all jet sources.

In radio-quiet quasars, usually a weak compact core is found. This
suggests that there is still an active jet, albeit one of lower power.
A typical quenching factor for the radio emission of radio-quiet jets
is $\sim 100$ \citep[e.g.,][]{KellermannSramekSchmidt1989}.  If we
assume that the radio emission is quenched by a factor 100, the jet
powers in in radio-quiet sources (RQQ) are reduced by a factor 30
(assuming $L_{radio} \propto P_{jet}^{1.4}$, e.g.,
\citealt{KoerdingFenderMigliari2006}).  We can compare the effect of
the RQQ in the KLF to that of RLQs and LLAGN-like jet sources.  As the
fraction of radio-loud sources is small at high luminosities (a few
percent, Fig.~\ref{Figfrac}), the number of radio-quiet sources is
approximately the number of all quasars.

\begin{figure}
\resizebox{8.4cm}{!}{\includegraphics{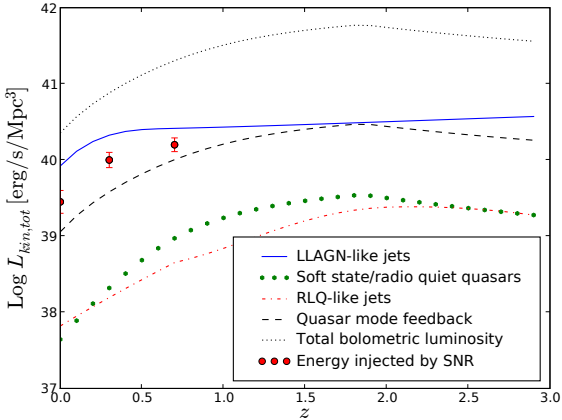}}
\caption{Total kinetic power injected into the ISM at a given redshift
  by AGN. We show the integrated kinetic energy injected into the IGM
  by LLAGN, by radio loud quasars and by radio quiet quasars as well
  as the power available for feedback during the quasar phase (e.g.,
  as radiative feedback).  The points denote the kinetic energy
  injected by supernovae. Additionally, we show the total bolometric luminosity, i.e. the total power liberated through accretion. \label{FigKinEvo}}
\end{figure}
The resulting jet KLF is shown in the left-hand panel of
Fig.~\ref{FigLJet}.  The LLAGN-like jet sources inject significantly
more kinetic energy into the ISM than both radio-loud and radio-quiet
quasars. The right-hand panel shows that this is true at all
redshifts. This can be seen even better in Fig~\ref{FigKinEvo}, where
we plot the total jet power integrated over all accretion rates as a
function of redshift.  The LLAGN-like jet sources dominate the total
kinetic jet power output by an order of magnitude. The contribution of
the RLQ-like jet sources is just comparable to those of radio quiet
sources. RLQ create jets that are roughly two orders of magnitude
stronger than those of the RQQ, but only a few percent of quasars are
radio-loud (RLQ-like jet sources). Both factors average out and we
find that the contribution of the sample of RQQ and RLQ to the kinetic
output is roughly similar.

In addition to the kinetic power injected by AGN, we show the energy
available for ``quasar-mode'' feedback in Fig.~\ref{FigKinEvo},
assuming that 5 $\%$ of the quasar's bolometric luminosity is
available to heat the surrounding gas \citep[see][\emph{e.g.}; we
  discuss the possible nature of the coupling between gas and
  radiation in
  sect.~\protect\ref{s:disc.feedback}]{diMatteoSpringelHernquist2005}.
Even including quasar-mode feedback, the population of low-luminosity
AGN with their LLAGN-like jets dominates the total power available for
feedback at low redshifts. As the cosmological evolution of the
high-luminosity quasars is stronger than for the low-luminosity end,
radiative feedback increases in importance until it reaches the same
order of magnitude as kinetic feedback at a redshift close to 2. We
note that it is not yet possible to \emph{observe} an LLAGN ARF at
$z=1$--$2$.  The only hints to the evolution of the low-luminosity
radio LF come from the fact that the LLAGN ARF smoothly extends the
BLF to lower luminosities, which suggests that one can use evolution
of the low-luminosity end of the BLF to estimate the evolution of all
LLAGN.

It is now widely accepted that the kinetic energy input from
supernovae has a profound impact on the formation and evolution of
galaxies \citep[see][e.g., and references therein]{Efstathiou2000}. A
typical supernova injects around $10^{51}$\,erg of kinetic energy into
the ambient medium
\citep[e.g.,][]{Chevalier1977,KorpiBrandenburgShukurov1999}. Assuming
that both core-collapse and type Ia supernovae create a kinetic energy
of $10^{51}$ erg, we can estimate the kinetic power injected into the
interstellar medium by supernovae, assuming the high-redshift
supernova rates estimated from the ``Great Observatories Origins Deep
Survey'' by \citet{DahlenStrolgerRiess2004}, and compare it to the
power created by jets of supermassive black holes. In
Fig.~\ref{FigKinEvo} we present our power estimate of jets, radiative
processes and supernovae. While the contributions of high-power jets
(i.e., radio-loud quasars) and radiative processes to the heating of the
ambient medium are significantly below the supernova heating
rate, LLAGN-like jet sources inject a similar amount of power into the
interstellar/intergalactic medium as supernovae. Also the jets of
stellar accreting black holes and neutron stars contribute to the
kinetic power, albeit at lower rates
\cite[e.g.,][]{FenderMaccaronevanKesteren2005}.

Of course, the impact of jets on the galaxy formation process depends
critically on the relative magnitude of the jet power, the duration of
each jetted phase, the collimation and absolute jet power (see section \ref{s:disc.feedback}), 
and the binding energy and cooling time of gas in
the galaxy.  Therefore, it would be highly desirable to derive
Fig.~\ref{FigKinEvo} separately for different galaxy types,
luminosities and black hole masses.  This requires knowledge of the
radio and bolometric luminosity functions separately for each
subcategory, as well as reaching lower luminosities at higher
redshifts than is presently possible, and is beyond the scope of this
work.

\section{Discussion}
\label{sediscussion}

\subsection{The break in the BLF and the missing LLAGN} 
\label{s:disc.break}

In Sect.~\ref{selocallum} we showed that the low-luminosity radio ARF
is roughly in agreement with the AGN BLF. However, supermassive black
holes at low accretion rate are likely to be radiatively inefficient
and would not be classified as a quasar.  Assuming the usual high
radiative efficiency of quasars will then underestimate the accretion
rate. At the lowest luminosities, this suggests that the radio ARF and
the BLF should increasingly deviate from each other with decreasing
accretion rates, with the radio ARF exceeding the AGN BLF
significantly. This is not what we find in the ARFs we have derived
(Figures~\ref{FigLumZ0} and \ref{SlowFast}). Here, we explain the observed break in the BLF as due to the onset of inefficient accretion and the ``missing'' radio LLAGN as due to the fact that there is only a finite number of supermassive BHs.

We have several constraints for modeling the accretion rate function
of supermassive black holes. First, the local black-hole mass function
has been estimated by \citet{ShankarSalucciGranato2004}. We can
therefore write the accretion rate function as:
\begin{equation}
\Phi(\dot{M}) = \int \Phi_{\mathrm{BH}}(M) \bar{\Phi}(M,\dot{M}) d\log M,
\label{eqPhiMdot}
\end{equation}
where $\Phi(\dot{M})$ is the measured accretion rate function in units
of Mpc$^{-3}\ d\log_{10}\dot{M}^{-1}$. Secondly, as the number of AGN
integrated over all accretion rates down to 0 has to be equal to the
number of all supermassive black holes, it follows:
\begin{equation}
1 = \int \bar{\Phi}(M,\dot{M}) d\log \dot{M}. \label{eqnorm}
\end{equation}
The distribution $\bar{\Phi}(M,\dot{M})$ describes the probability to
find a black hole of mass $M$ with an accretion rate $\dot{M}$. In
principle that function can be an arbitrary function and there is no
unique inversion that obtains $\bar{\Phi}(M,\dot{M})$ from
eq.~\ref{eqPhiMdot} with only the knowledge of $\Phi(\dot{M})$ and
$\Phi_{\mathrm{BH}}(M)$. As a first approach, let us assume that the
distribution depends only on $\dot{m}$, the accretion rate in
Eddington units. This assumption is equivalent with a distribution
$\bar{\Phi}(M,\dot{M})$ that separates into a function depending only
on $\dot{M}$ and another one depending only on $M$:
$\bar{\Phi}(M,\dot{M}) = \Phi_1(M)\Phi_2(\dot{M})$.  In astrophysical
terms, this assumption means that the Eddington luminosity,
\emph{i.e.}, the black-hole mass, sets the total power that is
available for feedback from a given black-hole mass, independently of
the host galaxy properties.

It is generally assumed that strongly accreting objects are
radiatively efficient.  As the BLF is well-described by a power law at
high luminosities (above the break), it is likely that the
distribution $\bar{\Phi}(\dot{m})$ is a power law at high Eddington
ratios. However, as the power-law index of the luminosity function is
steep ($\sim -2$), it has to have a cut-off towards low luminosities
so that its integral does not diverge. As there are known strongly
sub-Eddington sources (e.g., Sgr A$^{*}$), a hard cut-off would not be
appropriate, and we use a broken power law of the form:
\begin{equation}
 \bar{\Phi}(\dot{m}) \propto \left\{\begin{array}{ll}
 \dot{m}^{\gamma_2} & \mbox{for}\ \dot{m}_{\mathrm{break}} \leq \dot{m}
 \\ \left(\dot{m}/\dot{m}_{\mathrm{break}}\right)^{\gamma_1}
 \dot{m}^{\gamma_2}_{\mathrm{break}} & \mbox{for} \ \dot{m} \leq
 \dot{m}_{\mathrm{break}}.
 \end{array} \right. \label{eqmdotdist}
\end{equation}
As the function has to be integrable with respect to $d\log\dot{m}$ we
require $\gamma_1 \geq 0$. At high accretion rates, we truncate the
power law at 10 Eddington rates. The distribution will be normalized
according to eq.~(\ref{eqnorm}).

\begin{figure}
\resizebox{8.4cm}{!}{\includegraphics{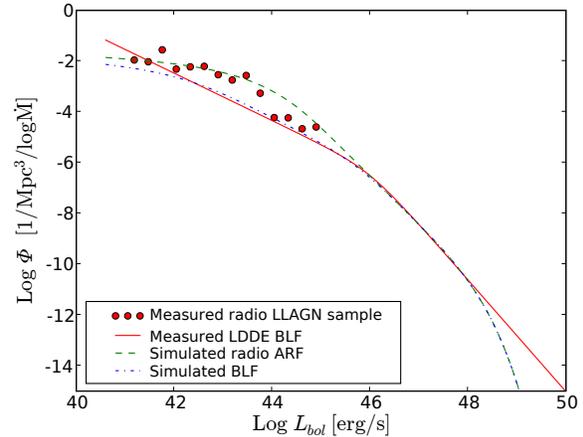}}
\caption{Measured ARF and BLF together with our model for the
  LFs. While the model for the ARF assumes that the accretion rate is
  directly measured we assume inefficient accretion below a critical accretion
  rate for the simulated BLF. We can reproduce both, the ARF and the BLF with
  our simple model. \label{fiARFQLF}}
\end{figure}
While the ARF obtained from the radio estimates should directly
measure $\Phi(\dot{M})$, the bolometric luminosity depends linearly on
the accretion rate only for the high-accretion rate sources. For
strongly sub-Eddington sources we have to translate the accretion rate
to a bolometric luminosity.  The analogy of AGN to XRBs suggests that
the bolometric luminosity of an accreting BH can be described as:
\begin{equation}
L_{\mathrm{bol}} = \left\{ \begin{array}{ll} L_{\mathrm{Edd}}\;\dot{m}
  & \mathrm{for}\ \dot{m} \geq \dot{m}_{\mathrm{crit}} \\ 
  L_{\mathrm{Edd}}\; \dot{m}_{\mathrm{crit}}
  \left(\frac{\dot{m}}{\dot{m}_{\mathrm{crit}}}\right)^2 &\mathrm{for} \ \dot{m} <
  \dot{m}_{\mathrm{crit}} \end{array} \right. \label{eqbol}
\end{equation}
where $\dot{m}$ is the accretion rate in Eddington units and
$\dot{m}_{\mathrm{crit}} \approx 0.05$. Note that the
$\dot{m}_{\mathrm{break}}$ used in the previous paragraph has no
direct relation to $\dot{m}_{\mathrm{crit}}$.  The exact value of
$\dot{m}_{\mathrm{crit}}$ seems to be different for the rise of each
XRB outburst $(0.03 \leq \dot{m}_{\mathrm{crit}} \leq 0.3)$. On the
decline it seems to be more stable $(\dot{m}_{\mathrm{crit}} \sim
0.02)$. For our simple model we will use a constant critical accretion
rate of $\dot{m}_{\mathrm{crit}} \sim 0.05$.

We can now combine the measured supermassive black hole function
(eq.~\ref{eqBHfunc}) with our assumptions about the accretion rate
distribution (eq.~\ref{eqmdotdist}) to obtain an accretion rate
function, $\Phi(\dot{M})$. This function can be compared directly to
the ARFs obtained from the radio emission. To obtain a BLF we include
the description given in eq.~\ref{eqbol} to translate from accretion
rates to luminosities for sources of a given mass. The free parameters
are $\gamma_{1,2}$ and the break ($\dot{m}_{\mathrm{break}}$) in the
power-law distribution of $\dot{m}$, and to some extent the critical
accretion rate $\dot{m}_{\mathrm{crit}}$. The latter is constrained
from the observations of XRBs and the finding that this value seems to
be similar in AGN \citep{Jester2005,KoerdingFenderMigliari2006}.

In this paper we only try to explain the small difference between the
measured radio ARF and the BLF at low luminosities. Thus, we will not
explicitly find a best-fit model to the data, but only compare a
plausible model to the ARF and the BLF.

The high-luminosity slope of the ARF and the BLF is mainly affected by
the power-law index $\gamma_2$ at high accretion rates. The position
of the break in the accretion rate distribution
($\dot{m}_{\mathrm{break}}$) strongly affects the normalization of the
BLF. If the break is too low, most BHs have a very low $\dot{m}$ and
there are not enough bright quasars. Thus, both $\gamma_2$ and
$\dot{m}_{\mathrm{break}}$ can be estimated from the high-luminosity
end of the BLF above the break. The low-luminosity index $\gamma_1$
has a rather minor effect on both the ARF and the BLF as long as
$\gamma_1 > 0$.  In Fig.~\ref{fiARFQLF} we show the local radio ARF
and the BLF together with one realization of our model ($\gamma_2 =
-1.9$, $\dot{m}_{\mathrm{break}} = 2\times 10^{-3}$, and $\gamma_1 =
1$). Both the BLF and the radio ARF are reproduced.

The features in the BLF and ARF can be understood as follows.  The
break in the BLF is created by radiative inefficiency at low accretion
rates (this effect of radiative inefficiency on the luminosity
function of AGN was exploited by \citealt{Jester2005} to look for
inefficiency in AGN).  The flattening of the ARF at lower luminosity
is due to the finite number of supermassive black holes -- if the ARF
did not flatten, there would need to be more low-luminosity AGN, not
leaving enough black holes to account for the observed high-luminosity
(quasar) population.

With the simple assumption that the distribution function of accretion
rates $\bar{\Phi}(\dot{m})$ depends only on the accretion rate in
Eddington units and can be described with a broken power law we seems
to be able to reproduce the observed LFs. However, there are several
suggestions that the distribution does depend on the black hole mass
\citep[e.g.,][]{AllenDunnFabian2006}. If we relax our assumption that
the distribution function $\bar{\Phi}(M,\dot{M})$ separates into
distribution for the accretion rate and one for the black hole mass,
we have enormous freedom to choose the distribution. We tried several
functions, e.g., we explored if the break of the broken power law in
the $\dot{m}$ distribution can depend on the mass, but were not able
to improve the fits significantly above the simple model we presented.

\citet{HopkinsHernquistCox2006a} model the quasar luminosity function
by considering the birth rate of quasars with a given peak luminosity
and the subsequent evolution of the quasar light curve in their
merger-driven model. They explain the break in the LF as a
characteristic luminosity at which the birth rate peaks.
\citet{HopkinsHernquistCox2006} examine the evolution of the light
curves in more detail and show that quasar light curves vary
systematically with peak luminosity, i.e., an individual black hole's
accretion rate evolution depends on the peak accretion rate. In the
framework of these models, $\bar{\Phi}(\dot{m})$ will vary with
redshift depending on the age distribution and lightcurves of
currently active black holes, and possibly in different ways from the
variation we would require.  A detailed comparison between our and the
\citeauthor{HopkinsHernquistCox2006} models woudld require knowledge
of the high-redshift black-hole mass and LLAGN luminosity functions,
which are however not yet available.

We note that our model suggests that the weakest LLAGN do not only
have a small central BH but should also be fed at a strongly
sub-Eddington (and even sub-Bondi) accretion rate \citep[see, e.g.,
  the simulations and model-fitting for Sgr A$^*$
  []{YuanQuataertNarayan2003,CuadraNayakshinSpringel2006,MarroneMoranZhao2007}.

\subsection{Implications for feedback}
\label{s:disc.feedback}

\subsubsection{Available feedback power as function of accretion rate}

In sect.~\ref{sekinfeed} we have confirmed earlier findings that AGN
at lower luminosities (e.g., LINERs and low-luminosity FR-I RGs) are
likely to dominate the total kinetic power output in the local
universe with their LLAGN-like jets
\citep{BestKaiserHeckman2006,HeinzMerloniSchwab2007}. Their total
kinetic power is comparable to the power injected into the ISM by
supernovae. The population of jets in radio-loud and quiet quasars
carry more than an order of magnitude less power than the jets of
low-luminosity AGN; in fact their total power is not much greater than
that injected by jets of XRBs. At higher redshifts the power injected
into the IGM by quasars increases more strongly than the power
injected by the lower-luminosity objects. However, the low-luminosity
objects still seem to dominate.  If a fixed fraction of the luminosity
of quasars is available for feedback even in the absence of strong
jets (e.g. by direct radiative heating or via radiation
pressure-driven winds), feedback from luminous quasars also increases
with redshift.  For a feedback fraction of 5\%, this ``quasar mode''
would make a significant contribution to the total power available for
feedback at redshifts close to 2, and even dominate the total feedback
for fractions larger than 5\%. This is certainly only true when
summing over the entire population of all supermassive black holes; a
different scenario is likely to be found in the center of clusters and
in individual galaxies.  Again, we stress that radio and bolometric
luminosity extending to lower luminosities, higher redshifts, and
separated by host galaxy type, are needed to assess the impact of
feedback separately for galaxies of different galaxy type, luminosity,
or black hole mass.

Nevertheless, from the models used to explain the measured radio
accretion rate function, kinetic power function and the bolometric
luminosity, we \emph{can} deduce feedback properties of individual
objects. The analogy of AGN with XRBs and our explanation of the break
in the quasar LF in the preceding section suggest that all accreting
black holes below a critical accretion rate around $3 \times 10^{-2}$
of the Eddington rate are accreting inefficiently and in a
jet-dominated mode. Above that critical accretion rate, a source can
either be in the hard or hard intermediate state and continue to be
strongly jet emitting, or the jet power is quenched by a factor $\sim
30$. With the efficiencies discussed above, we obtain the following
approximations for the kinetic power available for feedback, as
function of the fundamental parameters $\dot{m}$ and
$L_{\mathrm{Edd}}$, \emph{i.e.}, $M_{\mathrm{BH}}$:
\begin{equation}
P_{\mathrm{kin}} = \left\{\begin{array}{ll} 
0.2\,\dot{m}\, L_{\mathrm{Edd}} & \mbox{if}\ \dot{m} \leq \dot{m}_{\mathrm{crit}} \\ 
0.2\,\epsilon_{\mathrm{FR-II}}\, \dot{m}\, L_{\mathrm{Edd}} &
\mbox{if}\ \dot{m} > \dot{m}_{\mathrm{crit}},\ \mbox{with jet}\\ 
0.007\,\dot{m}\,L_{\mathrm{Edd}} & \mbox{if}\ \dot{m} >
\dot{m}_{\mathrm{crit}},\ \mbox{no jet}
\end{array} \right.
\label{eq:KinFeedbackPower}
\end{equation}
The ``with jet'' and ``no jet'' cases refer to radio-quiet quasars and
radio-loud quasars/radio galaxies, respectively.  For the latter
category, it is unlikely that much of the power in the jet will be
deposited inside the galaxy, as essentially all the energy is
transported out into the lobes which can be located anything from tens
of kiloparsecs to a few Megaparsecs away from the central galaxy. The
additional efficiency factor $\epsilon_{\mathrm{FR-II}}$ accounts for
this effect. 

It is not yet certain if and how radio-quiet quasars actually provide
energy to heat the ambient medium. One possibility is that a
certain fraction of the bolometric luminosity couples to the gas
through line driving \citep[e.g.,][]{FabianCelottiErlund2006}. Another
possibility is that the quasar launches a wind that heats the gas
through its kinetic power. It is likely that this process is also
linearly coupled to the bolometric luminosity.  According to the model
used in this paper, low accretion-rate objects are inefficiently
accreting up to the critical accretion rate. Accretion above that rate
is radiatively efficient. The approximate dependence of the bolometric
luminosity on accretion rate is given in eq.~(\ref{eqbol}). It has
been argued that roughly 5$\%$ of the bolometric luminosity is
available to heat the surrounding gas
\citep[e.g.,][]{diMatteoSpringelHernquist2005}. We will use this model
to compare the effects of radiative feedback to the kinetic heating
due to jets, i.e., the radiative feedback is given by
\begin{equation}
P_{\mathrm{radfeed}} = 0.05 L_{\mathrm{bol}}.
\label{eq:RadFeedbackPower}
\end{equation}

\begin{figure}
\resizebox{8.4cm}{!}{\includegraphics{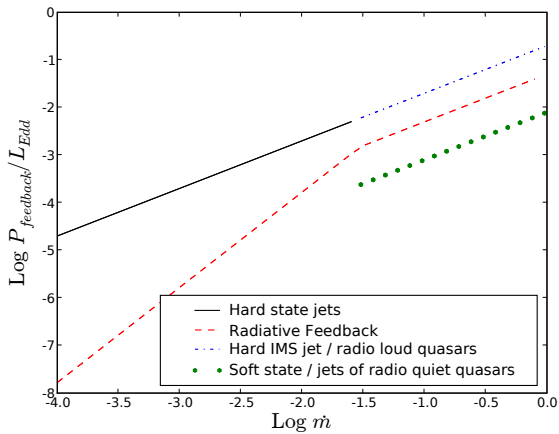}}
\caption{Power available for feedback as function of accretion rate,
  both in Eddington units (from eqns.~\ref{eq:KinFeedbackPower} and
  \ref{eq:RadFeedbackPower}). Sources are kinetically dominated at low
  accretion rates, but radiative feedback may dominate at high
  accretion rates. A simple linear dependency may correctly describe
  the total available power as function accretion rate within
  measurement uncertainties. \label{fifeedback}}
\end{figure}
Figure~\ref{fifeedback} shows the power available from a single source
(eqns.~\ref{eq:KinFeedbackPower} and \ref{eq:RadFeedbackPower}) via
both feedback mechanisms, radiative and kinetic (setting
$\epsilon_{\mathrm{FR-II}}=1$ for jet-active radio-loud quasars/radio
galaxies), as function of accretion rate. At low accretion rates, the
kinetic feedback of the jet dominates the total power output. At high
accretion rates, the radiative feedback is efficient and contributes
significantly to the total power. The jet is either quenched and the
power in the jet is negligible compared to the power available via
radiative feedback -- or the source is still jet-active and radiative
and kinetic feedback are comparable for our parameters. If most of the
energy in the jet is injected into distant lobes and not deposited
inside the galaxy and $\epsilon_{\mathrm{FR-II}} \ll 1$, as is very
likely the case, the kinetic feedback of radio-loud quasars is
actually much lower than the \emph{maximum available} feedback energy
shown in our figures.

At low accretion rates, the total power available for feedback is
linearly dependent on the accretion rate and dominated by the kinetic
power of the jets. At high accretion rates, the total power is either
dominated by radiative feedback, or a combination of radiative and
kinetic power. Again, the total feedback depends linearly on the
accretion rate with a proportionality factor similar to that in the
hard state. Thus, the assumption that the energy available for
feedback is roughly $P_{\mathrm{feedback}} \sim 0.01 \dot{M} c^2$,
seems to be correct within a factor 2 for all accretion rates; i.e.,
$P_{\mathrm{feedback}}$ is 10\% of $0.1 \dot{M} c^2$, the power
liberated by the accretion process. 

\subsubsection{Total jet feedback may exceed total supernova feedback}

As mentioned, \citet{BinneyOmma2007} have suggested recently that the
jet power estimates from bubbles underestimate the total jet power by
a factor of order 6. In this case, all jet lines in
Fig.~\ref{fifeedback} move up by a factor 6. The jet power would then
be an order of magnitude above the power available via radiative
feedback. Strongly jet emitting-sources (including both LLAGN and
FR-II radio galaxies and quasars) would have significantly more power
available for feedback than radio-quiet objects. As a population,
LLAGN-like jet sources would inject significantly more power into the
ISM than the kinetic power created by all supernovae.

\section{Summary and conclusion}

For both AGN and X-ray binaries, both the accretion rate and the jet
power can be estimated from either the core radio luminosity
\citep{KoerdingFenderMigliari2006} or the extended low-frequency radio
luminosity \citep{WillottRawlingsBlundell1999}. Using these accretion
rate estimates, we construct accretion rate functions (ARFs) of
jet-emitting sources. The luminosity function (LF) of low-luminosity
sources -- which likely have slow jets -- is roughly in agreement with
the bolometric luminosity function (BLF) of AGN (Fig.~\ref{FigLumZ0})
as determined by \citet{HopkinsRichardsHernquist2007}. The ARF of
RLQ-like jet sources is roughly 1\,dex below the BLF
(Fig.~\ref{Figfrac}).  This provides a measurement of the radio-loud
fraction with a direct physical meaning: the ratio of volume densities
of radio-loud and radio-quiet sources at a given accretion
rate. However, the exact value is unfortunately strongly dependent on
the normalizations used to derive accretion rates.

We have developed a simple model based on the universality of
accretion physics in XRBs and AGN that reproduces both our ARF as well
as the BLF. At low luminosities, all sources are radiatively
inefficient and the luminosity depends quadratically on the accretion
rate (eq.~\ref{eqbol}). Sources are radiatively efficient only at high
luminosities ($\ga 3\%$ Eddington). In our model, we assumed that
the distribution of accretion rates does not depend on the black-hole
mass and can be written as a simple broken power law for the
Eddington-scaled accretion rate (eq.~\ref{eqmdotdist}). With this
model we can reproduce the measured ARF as well as the local BLF
(Fig.~\ref{fiARFQLF}). Thus, this model can solve the problem that there seem to be too few LLAGN compared to the number of weak quasars. In this simple model the break in the
luminosity function is due to the inefficient accretion at low
accretion rates \citep[compare][]{Jester2005}.

Using the corresponding jet power measures, we have calculated kinetic
luminosity functions and their cosmological evolution. Our findings
support the idea that the majority of the kinetic power is created by
lower-luminosity AGN with their likely slow jets. The highly luminous
radio-loud quasars do not contribute significantly due to their low
number density. The total power injected into the ambient medium by
jets from AGN is comparable or even exceeds the effect of supernovae
(the situation may and will be different for individual objects). Only
at high redshifts, ``quasar-mode'' feedback mechanisms provide a
comparable amount of energy for heating the interstellar/intergalactic
gas. Finally, we discuss the effects of the different feedback
mechanisms and find that a roughly constant fraction (5-10$\%$) of the
accretion power is available for feedback, independent of the nature
of the AGN.

\section*{Acknowledgements}

We acknowledge helpful discussions with Rachel Somerville, Eric Bell
and Frank van den Bosch.  We are grateful to the anonymous referee for
helping us improve the presentation of this material.

\label{lastpage}

\end{document}